\documentclass[10pt,twocolumn,prl,aps,showpacs]{revtex4}

\usepackage{amsmath}
\usepackage{amssymb}
\usepackage{amsfonts}
\usepackage{graphicx}
\usepackage{dcolumn}
\usepackage{hyperref}

\def\3nab{\tilde{\nabla}}

\def\be {\begin{equation}}
\def\ee {\end{equation}}
\def\bea {\begin{eqnarray}}
\def\eea {\end{eqnarray}}

\newcommand{\beaa}{\begin{eqnarray*}}
\newcommand{\eeaa}{\end{eqnarray*}}

\newcommand{\e}{{\rm e}}
\def\case#1/#2{\textstyle\frac{#1}{#2}}
\usepackage{graphicx}
\usepackage{dcolumn}
\usepackage{bm}
\usepackage{longtable}

\def\cqg{{\em Class. Quantum Grav.\/} }
\def\grg{{\em Gen. Rel. Grav.\/} }
\def\prd{{\em Phys. Rev.\/} {\bf D}}

\def\aph{{\em Ann. Phys. (NY)\/} }
\def\plb{{\em Phys. Lett.\/} {\bf B}}

\begin{document}

\title{On the $\Lambda$CDM Universe in $f(R)$ gravity}

\author{Peter K. S. Dunsby$^{1,2,3}$, Emilio Elizalde$^{4}$,  Rituparno Goswami$^{1,2}$,  Sergei Odintsov$^{4,5}$ and Diego Saez-Gomez$^{4}$}

\affiliation{1. Centre for Astrophysics, Cosmology and Gravitation,
University of Cape Town, Rondebosch, 7701, South Africa}

\affiliation{2. Department of Mathematics and Applied Mathematics,
University of Cape Town, 7701 Rondebosch, Cape Town, South Africa}

\affiliation{3. South African Astronomical Observatory,
 Observatory 7925, Cape Town, South Africa.}

\affiliation{4.  Institut d'Estudis Espacials de Catalunya, ICE (IEEC-CSIC), Campus UAB,  Facultat Ci\`encies, Torre C5-Par-2a pl, E-08193 Bellaterra (Barcelona) Spain}

\affiliation{5. ICREA, Barcelona}

\date{\today}

\begin{abstract}
Several different explicit reconstructions of $f(R)$ gravity are obtained from the background FRW expansion history. It is shown that the only theory whose Lagrangian is a simple function of the Ricci scalar $R$, that admits an exact $\Lambda $CDM expansion history is standard General Relativity with a positive cosmological constant and the only way to obtain this behaviour of the scale factor for more general functions of $R$ is to add
additional degrees of freedom to the matter sector. 
\end{abstract}

\maketitle

\section{Introduction}
After more than one hundred years, General Relativity (GR) is still
considered to be the best fundamental theory for the description of
the gravitational action. When applied to cosmology, assuming
homogeneity and isotropy (encoded in the Friedmann-La\^{i}matre-Robertson-Walker 
metric), together with a fluid description of baryons, Cold Dark Matter 
(CDM) and radiation, GR gives rise to a set of field equations which 
when solved produce the simplest expanding cosmology - the 
Friedmann model, governing the dynamics of the cosmological scale-factor 
$a(t)$. This model has been remarkably successful, giving for example 
the correct light element abundances and explaining the origin 
of the Cosmic Microwave Background Radiation (CMBR). In the 
last two decades, however, advances in observational cosmology 
appears to suggest that if one wishes to retain the FLRW metric, the 
universe must have undergone two periods of accelerated expansion. 
The first period of acceleration, known as the inflationary epoch is 
needed to explain the flatness problem and the near-scale invariant spectrum 
of temperature fluctuations observed in the CMBR, while the second period 
explains the dimming of distant type Ia supernovae relative to Einstein-de Sitter 
universe model.  In order to explain these periods of acceleration, the strong energy
condition ($\rho+3p\geq 0$) needs to be violated. In the case of
inflation, this is achieved by introducing a dynamical scalar field,
while the present day acceleration is most easily explained with the
introduction of a cosmological constant. The resulting description of
the Universe, in which the recent expansion history is driven by a
cosmological constant and ordinary matter is dominated by a CDM
component has become known as the $\Lambda$CDM or {\em Concordance
Model} \cite{concordance}. Unfortunately, this beautifully simple
phenomenological model, which appears to fit all currently available
observations (Supernovae Ia \cite{sneIa}, CMBR anisotropies \cite{cmbr}, 
Large Scale Structure formation \cite{lss}, baryon oscillations \cite{bo} 
and weak lensing \cite{wl}) is affected by significant fine-tuning problems 
related to the vacuum energy scale and therefore it is important to investigate
alternatives to this description of the Universe.

Currently, one of the most popular alternatives to the {\em $\Lambda$
CDM} model is based on modifications of the standard Einstein-Hilbert
action. This is due to the fact that these changes naturally
admit a phase of late time accelerated expansion (an early Universe
inflationary phase is also possible \cite{star80}). In this way Dark
Energy can be thought of as have a geometrical origin, rather than be
due to the vacuum energy or additional scalar fields which are added
by hand to the energy momentum tensor. As a result, the cosmology and
astrophysics of modified gravity is currently an extremely active
area of research (see the recent reviews \cite{reviews} and
references therein).

One of the simplest extensions of GR is based on gravitational
actions which are non-linear in the Ricci curvature $R$ and$/$or
contain terms involving combinations of derivatives of $R$
\cite{DEfR, ccct-ijmpd,otha,cct-mnras,kerner, teyssandier,magnanoff}. 
An important feature of these theories is that the field equations can be written in a way
which makes it easy to compare with GR. This is done by moving all
the higher-order corrections to the curvature onto the RHS of the
field equations and defining an "effective" source term, often
described as the {\em curvature fluid}. Once this has been done, it
is easy to see that how a change of sign in the deceleration
parameter of the FLRW cosmology can occur, leading to a period of
late-time acceleration.

Studies of the physics of these theories is however hampered by
the complexity of the field equations, making it difficult to obtain
both exact and numerical solutions which can be compared with
observations. Theses problems can be reduced somewhat by using the
theory of dynamical systems \cite{DS}, which provides a relatively
simple method for obtaining exact solutions and a (qualitative)
description of the global dynamics of these models for a given $f(R)$
theory \cite{Dyn}.

Another useful approach is to assume that the expansion history of
the universe is known exactly, and to invert the field equations to
deduce what class of $f(R)$ theories give rise to this particular
cosmological evolution.

This has been done recently for exact power-law solutions for the
scale factor, corresponding to phases of cosmic evolution when the
energy density is dominated by a perfect fluid. It was found that
such expansion histories only exist for $R^n$ gravity \cite{julien}.
A more extensive analysis of {\em reconstruction methods} has been
carried out in \cite{odintsov} to obtain theories which give an
approximate description of deceleration-acceleration transitions in
cosmology and in \cite{reconstruction1} a powerful approach to reconstruction 
based on  standard cosmic parameters instead of a time law for the scale 
factor was introduced.

In this paper we perform a number of explicit reconstructions which
lead to a number of interesting results. We find, for example, that the
only real valued Lagrangian $f(R)$ that is able to mimic an exact
$\Lambda$CDM expansion history for a universe filled with dust-like
matter is the Einstein-Hilbert Lagrangian with positive cosmological
constant. This does not mean that $f(R)$ gravity is incompatible with
an exact $\Lambda$CDM expansion history. In fact we further show that
in a universe filled with a both a minimally-coupled non-interacting
massless scalar field and dust-like matter, a theory of gravity can
be found which exactly mimics the $\Lambda$CDM expansion history,
making it impossible to distinguish it from GR at the level of the FLRW
background. Moreover, number of realistic models may also mimic late-time 
acceleration epoch approximately.

\section{Field equations for homogeneous and isotropic spacetimes in
$f(R)$ gravity}

We consider the following action within the context of 4-dimensional
homogeneous and isotropic spacetimes, i.e., the FLRW universes
with negligible spatial curvature:
\begin{equation}\label{lagr f(R)}
\mathcal{A}=\frac12\int d^4 x \sqrt{-g}\left[f(R)+{\cal
L}_{m}\right]\;,
\end{equation}
where $R$ is the Ricci scalar, $f$ is general differentiable
(at least $C^2$) function of the Ricci scalar and $\mathcal{L}_m$
corresponds to the matter Lagrangian. Units are chosen so that $c=8\pi G=1$.

The field equations for homogeneous and isotropic spacetimes are the
{\em Raychaudhuri equation}
\bea
\label{ray}
3\dot{H}+3H^2&=&-\frac{1}{2f'}\left[\rho+3p+f-f'R+3H f''
\dot{R}\right.\nonumber\\
&&\left.+3f'''\dot{R}^2+3f''\ddot{R}\right]\;,
\eea
where $H$ is the Hubble parameter; 
the {\em Friedmann equation}
\begin{equation}\label{fried}
3H^2= \frac{1}{f'}\left[\rho+\frac{Rf'-f}{2}-3H f'' \dot{R}\right]\;;
\end{equation}
the {\em trace equation}
\be
\label{trace}
3\ddot{R}f''=\rho-3P + f'R-2f-9H f''\dot{R}-3f'''\dot{R}^2\;;
\ee
and the {\em energy conservation equation} for standard matter
\begin{equation}\label{cons:perfect}
\dot{\rho}=-\Theta\left(\rho+P\right)\;.
\end{equation}
Combining the Friedmann and Raychaudhuri equations, we obtain
\begin{equation}
R=6\dot{H}+12H^2\,,\label{R}
\end{equation}
which is the usual definition for the Ricci scalar for homogeneous
and isotropic flat FRW spacetimes.

The Raychaudhury equation can be obtained by adding the Friedmann equation 
with it's time derivative (using the energy conservation
equation and the definition of the Ricci scalar). Therefore any
solution of the Friedmann equation is automatically a solution to the Raychaudhuri 
equation and hence it is sufficient to solve the Friedmann equation to reconstruct 
the theory of gravity.

\section{Reconstruction of a $f(R)$ theory that admits an exact
$\Lambda$CDM model}

As we know, the present day observations suggest that the variation
of the Hubble parameter with the redshift is sufficiently 
well described by the relation
\be
H(z)=\sqrt{\frac{\rho_0}{3}(1+z)^3+\frac{\Lambda}{3}}\;,
\label{hz}
\ee
where $\rho_0\ge0$ is the matter density (which consists of the
observed and the cold dark matter) and $\Lambda$ is the cosmological
constant.
In what follows we try and construct theories (belonging to the class of f(R)
gravity), that exactly mimic the above expansion history.

From equation (\ref{hz}) we see that the time derivative of the
scale factor $a(t)$ can be given as
\be
\dot{a}=\sqrt{\frac{\rho_0}{3a}+\frac{\Lambda}{3}}\;.
\label{adot}
\ee
Here we have used the usual definition of the redshift $1/a=1+z$.
From the above equation we can immediately calculate the second 
derivative of the scale factor, which is given by
\be
\ddot{a}=\frac{1}{2}(\dot{a}^2)_{,a}=\frac{2\Lambda a^3-\rho_0}{6a^2}\;.
\label{addot}
\ee
We know for a flat FLRW universe, the Ricci scalar is defined by
\be
R=6\left(\frac{\ddot{a}}{a}+\frac{\dot{a}^2}{a^2}\right)\;.
\label{R1}
\ee
Using equations (\ref{adot}) and (\ref{addot}) in equation (\ref{R1})
we obtain the Ricci scalar in terms of the scale factor as
\be
R(a)=\frac{4\Lambda a^3+\rho_0}{a^3}\;.
\ee
We would like to emphasize here that till now we haven't a priori
assumed any specific theory of gravity. Equation (\ref{adot}) is obtained 
by observations while equation (\ref{R1}) is a purely geometrical result for flat
homogeneous and isotropic spacetimes, independent of the theory of gravity.

We can now invert equation (\ref{R1}), to write the scale factor in
terms of the Ricci scalar
\be
a(R)=\left(\frac{\rho_0}{R-4\Lambda}\right)^{(1/3)}\;.
\label{aR}
\ee
We note that, since the scale factor has to be real, we considered
only the real root of equation (\ref{R1}), discarding the other two complex conjugate
roots. Also the positivity of the scale factor implies that the Ricci scalar
reaches the value $4\Lambda$, asymptotically in an infinite time. From equation
(\ref{aR}) we can calculate the Hubble parameter and the time derivative of Ricci 
scalar in terms of the Ricci scalar and these are given by
\bea
H(R)&=&\frac{1}{a(R)}\sqrt{\frac{\rho_0}{3a(R)}+\frac{\Lambda}{3}}\;,
\\
\dot{R}&=&R_{,a}(a(R))\sqrt{\frac{\rho_0}{3a(R)}+\frac{\Lambda}{3}}\;.
\eea
Similarly, using the energy conservation equation we can write the
matter density of the universe in terms of scale factor `$a$', or alternatively
$\rho(a(R))$.

To investigate which functions $f(R)$ exactly mimic the $\Lambda$CDM
expansion history, we substitute all the above quantities
written as a function of the Ricci scalar into the Friedmann equation, obtaining
\bea
-3(R-3\Lambda)(R-4\Lambda)f''(R)&&\nonumber\\
+\left(\frac{R}{2}-3\Lambda\right)f'(R)+\frac12f(R)-\rho(R)&=&0\;.
\label{fR1}
\eea
Since this equation has to be satisfied for all times (which imply
for all $R\ge4\Lambda\in\Re$), this becomes a differential equation for the
function $f(R)$ in $R$-space. It is easy to see that the above equation is an
exact inhomogeneous hypergeometric equation for the variable 
$x\equiv -3+R/\Lambda$, with $\rho(R)$ as the inhomogeneous term. 
The solution to the homogeneous part is given by
\bea
f(x)&=&C_1F\left([\alpha_+,\alpha_-],-\frac12;x\right)\nonumber\\
&&+ C_2x^{3/2}F\left([\beta_+,\beta_-],\frac52;x\right)\;,
\eea
where $\alpha_{\pm}=(-7\pm\sqrt{73})/12$,
$\beta_{\pm}=(-11\pm\sqrt{73})/12$ and $C_{1,2}$
are arbitrary constants of integration.

Let us now analyze the solution carefully. The two finite poles of
the hypergeometric equation are at $R=3\Lambda$ and $R=4\Lambda$ 
respectively. Since the allowed range of $R$ is $R\ge4\Lambda$, one of 
the pole is out of the range and the other is at the boundary. However we see that in
this range $x\ge1$. We know that the convergence of a hypergeometric
function for the variable `$x$' is guaranteed if $|x|<1$, and otherwise the function
is either divergent or complex valued. Indeed one can check explicitly that both the solutions
of the homogeneous equation are complex valued for $R\ge4\Lambda$.
Hence to ensure a real valued function $f(R)$, we must choose both the arbitrary
constants $C_{1,2}$ to be zero. This is interesting as it shows that there cannot be any real
valued function of Ricci scalar that can mimic a $\Lambda$CDM expansion for a vacuum
universe.

Therefore from now on we only consider the particular solution for the given 
inhomogeneous term $\rho(R)$. Let us now explicitly reconstruct the
theories for which a given matter field would obey a $\Lambda$CDM expansion history. 
We would like to note here that a different reconstruction approach related with local tests in
modified gravity was considered in ~\cite{olmo}.

\subsection{Reconstruction for dust-like matter}

Supposing the Universe is filled with dust-like matter ($w=0$). From
the energy conservation equation we have
\be
\rho(a)=\frac{\rho_0}{a^3}\;\Rightarrow\;\rho(R)=R-4\Lambda\;.
\ee
Substituting in the Friedmann equation (\ref{fR1}), we get the
particular solution as
\be
f(R)=R-2\Lambda \;,
\ee
which is the well known Lagrangian for general relativity with a
cosmological constant. This result is interesting as it proves that the only real valued
Lagrangian $f(R)$, that can mimic an exact $\Lambda$CDM expansion
history for a universe filled with dust-like matter, is the
Einstein-Hilbert Lagrangian with positive cosmological constant.

It is also important to note, however, that this is not the case, if
we put $\Lambda=0$. In that case the general solution of the Friedmann equation is
\be
f(R)=R+C_1R^{\alpha_+}+C_2R^{\alpha_-}
\ee
We see that the function is real valued for positive Ricci scalar and
hence there exist classes of real valued function $f(R)$, other than
GR, that can mimic a dust-like expansion history without the 
cosmological constant. But even a very small value of the cosmological 
constant would break this degeneracy, and in that case the theory must be GR.

\subsection{Reconstruction for perfect fluid with equation of state $p=-1/3\rho$}

A perfect fluid with an equation of state $p=-\frac13\rho$ is
physically interesting as it lies in the boundary of the set of matter fields that obey the {\it
strong energy condition}. In GR such fluids give rise to a {\it Milne Universe} which is a
{\it coasting} universe, and the Ricci scalar is proportional to the square of the
Hubble parameter. However as we would see, even this kind of fluids can also mimic a
$\Lambda$CDM Universe in higher order gravity.

Using the equation of state in the energy conservation equation and
supposing the present density of the fluid is $\rho_f$ we get
\be
\rho(a)=\frac{\rho_f}{a^2}\Rightarrow\rho(R)=[\rho_f(R-4\Lambda)]^{2/3}\;.
\ee
Now solving the Friedmann equation (\ref{fR1}), the particular solution is
\be
f(R)=\mu(R-4\Lambda)^{2/3}\;,
\ee
where $\mu$ is a constant depending on $\rho_f$.

\subsection{Reconstruction for multi-fluids}

Let us now consider that along with dust-like matter, a non-interacting stiff fluid is also present in the universe and their present densities are $\rho_0$ and $\rho_s$ respectively. This scenario is also possible if we have 
non-interacting minimally coupled massless scalar field with the dust-like matter. In this case,
from the conservation equation, total matter density is
\be
\rho(a)=\frac{\rho_0}{a^3}+\frac{\rho_s}{a^6}\Rightarrow\rho(R)=(R-4\Lambda)
+\frac{\rho_s}{\rho_0^2}(R-4\Lambda)^2\;.
\ee
Substituting this into the Friedmann equation (\ref{fR1}), we get the following
particular solution:
\be
f(R)=\mu_1R+\mu_2R^2+\mu_3\;,
\ee
where $\mu_n(n=1..3)$ are constants depending on $\rho_0$, $\rho_s$
and $\Lambda$. We therefore conclude that if the universe is filled with minimally coupled non interacting massless scalar field with dust-like matter, then the theory of gravity described above would exactly mimic a
$\Lambda$CDM expansion history and it is impossible to distinguish this from GR with present
cosmological observations for the background FLRW level.

\subsection{Reconstruction for non isentropic perfect fluids}

Non-isentropic perfect fluids are generally described by the equation
of state
\be
p=h(\rho,a)\;.
\ee
Using the energy conservation relation we get the required differential equation for
$\rho$ as
\be
\rho(a)_{,a}=-\frac{3}{a}(\rho+h(\rho,a))\;.
\label{eqnrho}
\ee
In general the above equation may not admit a closed form solution.
However the calculation gets much simpler if $h(\rho,a)$ is a separable function
of the form
\be
h(\rho,a)=w(a)\rho\;,
\ee
that is, the barotropic index of the fluid changes with time. In this
case we can directly integrate equation (\ref{eqnrho}) to get
\be
\rho(a)=\exp\left[-3\int \frac{1+w(a)}{a} da\right]\;.
\label{eqnrho1}
\ee
We then substitute this into the Friedmann equation to get the required
theory of gravity.

As a specific illustration, lets assume that the time dependent
barotropic index is given by
\be
w(a)=\frac{2\gamma-\nu a^3}{\gamma+\nu a^3}\;,
\ee
where $\gamma$ and $\nu$ are constants. As we can see in early times
the fluid has positive pressure and at late times this behaves like a cosmological constant.
Using (\ref{eqnrho1}) and (\ref{aR}), we get $\rho$ in terms of the Ricci scalar as
\be
\rho(R)=\frac{(\nu\rho_0+\gamma R-4\Lambda\gamma)^3}{\rho_0^2}\;.
\ee
Substituting this into the Friedmann equation (\ref{fR1}), we obtain the
particular solution to be
\be
f(R)=\mu_1R+\mu_2R^2+\mu_3R^3+\mu_4\;,
\ee
where $\mu_n(n=1..4)$ are constants depending on $\rho_0$, $\gamma$,
$\nu$ and $\Lambda$.

Let us consider another form of non-isentropic perfect fluids where
the equation of state is given by
\be
p=w\rho+h(a)
\ee
This equation of state is physically interesting, due to the presence
of the tuning function $h(a)$, which can compensate the effect of the 
fourth order gravity at intermediate times. In this way we can have a matter 
dominated Friedmann like epoch (necessary for structure formations in the Universe), 
followed by an accelerated expanding phase.

Integrating the energy conservation equation (\ref{eqnrho}), we obtain
\be
\rho(a)=\left[-\int 3a^{(2+3w)}h(a)da +C_1\right]a^{-3(1+w)}\;,
\ee
where $C_1$ is an arbitrary constant of integration. Thus, we see
that the effective density is that of an isentropic perfect fluid together
with a time dependent cosmological term. We can always tune this term 
to obtain a fit with cosmological observations.

As a specific example let us consider $h(a)\equiv a^{-12}$ and $w=0$.
This then implies that the matter field in the Universe is dust along with a time dependent
cosmological term which diverges at the {\it Big Bang singularity} and goes to zero 
sufficiently fast at the later epochs.

Integrating the energy conservation equation we obtain
\be
\rho(a)=\frac{\rho_0}{a^3}+\frac{\rho_1}{a^{12}}\;\Rightarrow\;\rho(R)=R-4\Lambda
+\frac{\rho_1}{\rho_0}(R-4\Lambda)^4\;.
\ee
Solving the Friedmann equation, we obtain the following particular solutions 
\be
f(R)=\mu_1R+\mu_2R^2+\mu_3R^3+\mu_4R^4+\mu_5\;,
\ee
where $\mu_n~(n=1..5)$ are constants depending on $\rho_0$, $\rho_1$
and $\Lambda$.

\section{Reconstruction of approximate $\Lambda$CDM models }

Let us now consider a useful technique proposed in Ref.\cite{NOSG},
where the Friedmann equations are written as functions of the number
of e-foldings instead of the time, $N=\ln \frac{a}{a_0}$. In such a
case, the Hubble parameter is given in terms of $N$ as:
\be
\label{S1}
H=g(N) = g \left(- \ln\left(1+z\right)\right)\ .
\ee
Then, the first FLRW equation for a flat universe yields,
\bea
\label{S2}
0 &=& -9 G\left(N\left(R\right)\right)\left(4
G'\left(N\left(R\right)\right)
+ G''\left(N\left(R\right)\right)\right) \frac{d^2
f(R)}{dR^2}\nonumber\\
&&+ \left( 3 G\left(N\left(R\right)\right)
+ \frac{3}{2} G'\left(N\left(R\right)\right) \right) \frac{df(R)}{dR}
\nonumber\\
&& - \frac{f(R)}{2}
+ \sum_i \rho_{i0} a_0^{-3(1+w_i)} \e^{-3(1+w_i)N(R)}\;,
\eea
where $G(N) \equiv g\left(N\right)^2 = H^2$, and the Ricci scalar
can be related with $N$ via $R=12G(N)+3G'(N)$. Then the equation (\ref{S2}) 
can be resolved  for a given Hubble expansion rate, and the corresponding $f(R)$ 
theory is reconstructed.

\subsection{Cosmological solutions in $f(R)$ gravity with the
presence of an inhomogeneous EoS fluid}

We consider now a Universe governed by some specific $f(R)$ theory in
the presence of a perfect fluid, whose equation of state is given by,
\be
p=w(a)\rho+\zeta(a)\;,
\label{S14}
\ee
where $w(a)$ and $\zeta(a)$ are functions of the scale factor $a$,
which could correspond to the dynamical behavior of the fluid and to
its viscosity. Let us write the FLRW equations for f(R) as following,
\bea
H^2=\frac{1}{3}(\rho'+\rho_{f(R)})\ ;, \\
2\dot{H}+3H^2=-(p'+p_{f(R)})\;,
\label{S15}
\eea
where $\rho'=\frac{\rho}{f'(R)}$ and $p'=\frac{p}{f'(R)}$. The
pressure and energy density with the subscript $f(R)$ contains the
terms corresponded to $f(R)$ and are defined as
\bea
\rho_{f(R)}=\frac{1}{f'}\left(\frac{Rf'-f}{2}-3H\dot{R}f''\right)\;,
\\
p_{f(R)}=\frac{1}{f'}\left(\dot{R}^2f'''+2H\dot{R}f''+\ddot{R}f''+\frac{1}{2
}(f-Rf')\right)\;.
\label{S16}
\eea
Then, by combining both FLRW equations, and using the equation of state defined in
(\ref{S14}), we can write
\be
\zeta(a)=\left(w(a)\rho_{f(R)}-p_{f(R)}-2\dot{H}-3(1+w(a))H^2\right)f'(R(a))\;.
\label{S17}
\ee
As $\zeta(a)$ just depends on the Hubble parameter and its
derivatives, for some specific solutions, any kind of cosmology can
be reproduced. Let us consider the example,
\be
\frac{3}{\kappa^2}H^2=G_{\rho}a^{-c}+G_{q}a^{d}\;,
\label{S18}
\ee
where $G_{\rho}$ and $G_q$ are constants. We can check that in this
solution, the first term in the r.h.s. corresponds to a fluid with
EoS $w_p=-1+c/3>-1$, while the second term, it has an equation of state
$w_q=-1-d/3<-1$, which corresponds to a phantom fluid. We could
consider a viable $f(R)$ function proposed in
\cite{F(R)UnfInfCosAcce3}, which is given by
\be
F(R)=\frac{\alpha R^{m+l}-\beta R^n}{1+\gamma R^l}.
\label{S19}
\ee
This function is known to pass the local gravity tests and could
contribute to drive the Universe to an accelerated phase. Then, by
introducing (\ref{S18}) and (\ref{S19}) in the expression for
$\zeta(a)$ in (\ref{S17}), we obtain the equation of state for the inhomogeneous
fluid that, together with $f(R)$, reproduces the solution
(\ref{S18}), which for $c=3$ and $d\geq0$ reproduces the $\Lambda$CDM
model, and probably drives the Universe evolution into a phantom
phase in the near future.

\section{Conclusion}

In this paper we have extended a the reconstruction programme for
$f(R)$ gravity obtaining several interesting explicit
reconstructions. In particular we find that the only real valued
Lagrangian $f(R)$ that is able to mimic an exact $\Lambda$CDM
expansion history for a universe filled with dust-like matter is the
Einstein-Hilbert Lagrangian with positive cosmological constant. This
does not mean that $f(R)$ gravity is incompatible with an exact
$\Lambda$CDM expansion history, only that one has to extend the
theory somewhat in order that this possibility can be realized. For
example in a universe filled with a both a minimally-coupled
non-interacting massless scalar field and dust-like matter, a theory
of gravity can be found which exactly mimics the $\Lambda$CDM
expansion history, making it impossible to distinguish it from GR
using measurements of the background cosmological parameters.
It is then an interesting problem to probe, how the perturbations 
in these modified theories can break this degeneracy, by predicting 
different structure formations, growth factor or cosmological gravitational waves 
from GR, that can be experimentally verified ~\cite{pert}. In fact the 
matter power spectrum for power law $f(R)$ theories satisfies the requirement 
for scale invariance and has distinct features that can be detected by
combining future cosmic microwave background (CMB) and large scale surveys (LSS)
data.  
Moreover, it remains the number of realistic modified gravities which
reproduce this late-time cosmic acceleration era approximately, where
essentially any degree of accuracy may be achieved. In that case, it is then 
required to have local constraints, like the Solar System or Post 
Newtonian constraints (see for e.g ~\cite{olmo} and the references therein, which rules out 
the Lagrangians which grow indefinitely as the Ricci scalar vanishes) 
to have a modified gravity theory that works for both local and cosmological scales. 

\acknowledgments
This work was supported by a MICINN (Spain)-NRF (South Africa) exchange project grant.

\end{document}